\documentclass{article}
\usepackage{dpdfd}

\usepackage{times}
\usepackage{soul}
\usepackage{url}
\usepackage[hidelinks]{hyperref}
\usepackage[utf8]{inputenc}
\usepackage[small]{caption}
\usepackage{graphicx}
\usepackage{amsmath}
\usepackage{amsthm}
\usepackage{booktabs}
\usepackage{algorithm}
\usepackage{algorithmic}
\usepackage[switch]{lineno}
\usepackage{utfsym}
\usepackage{amssymb}
\usepackage{threeparttable}
\usepackage{multicol}
\usepackage{multirow}
\newcommand{\myPara}[1]{\vspace{.05in}\noindent\textbf{#1}}
\newtheorem{definition}{Definition}
\newtheorem{lemma}{Lemma}
\graphicspath{{./figures/}}

\newtheorem{theorem}{Theorem}

\title{Model Conversion via Differentially Private Data-Free Distillation}

\author{
    Bochao Liu$^{1,2}$\and
    Pengju Wang$^{1,2}$\and
    Shikun Li$^{1,2}$\and
    Dan Zeng$^3$\and
    Shiming Ge$^{1,2}$\footnote{Shiming Ge is the corresponding author (geshiming@iie.ac.cn).}\\
    \affiliations
    $^1$Institute of Information Engineering, Chinese Academy of Sciences, Beijing 100085, China\\
    $^2$School of Cyber Security, University of Chinese Academy of Sciences, Beijing 100049, China\\
    $^3$School of Communication and Information Engineering, Shanghai University, Shanghai 200444, China\\
    \emails
    \{liubochao,~wangpengju,~lishikun,~geshiming\}@iie.ac.cn,~dzeng@shu.edu.cn\\
}

\begin{document}

\maketitle

\begin{abstract}
    While massive valuable deep models trained on large-scale data have been released to facilitate the artificial intelligence community, they may encounter attacks in deployment which leads to privacy leakage of training data. In this work, we propose a learning approach termed differentially private data-free distillation (DPDFD) for model conversion that can convert a pretrained model (teacher) into its privacy-preserving counterpart (student) via an intermediate generator without access to training data. The learning collaborates three parties in a unified way. First, massive synthetic data are generated with the generator. Then, they are fed into the teacher and student to compute differentially private gradients by normalizing the gradients and adding noise before performing descent. Finally, the student is updated with these differentially private gradients and the generator is updated by taking the student as a fixed discriminator in an alternate manner. In addition to a privacy-preserving student, the generator can generate synthetic data in a differentially private way for other downstream tasks. We theoretically prove that our approach can guarantee differential privacy and well convergence. Extensive experiments clearly demonstrate that our approach significantly outperform other differentially private generative approaches.
\end{abstract}

\section{Introduction}
    The success of deep neural networks in a wide array of applications \cite{jia2019efficient,jia2019towards,ye2020reinforcement} greatly owes to the open source of massive models. However, a major problem is that the training data of these models often contain a large amount of sensitive information that can be easily recovered with a few access to the models~\cite{fredrikson2015ccs,yang2019ccs}. \textit{How to protect such private information while maintaining the model performance} has attracted a lot of attentions. Differential privacy~(DP)~\cite{dwork2006calibrating} is a common technique for protecting privacy. \cite{abadi2016deep} guaranteed that the model was differentially private regarding the training data by clipping gradients and adding Gaussian noise to gradients. However, the model accuracy decreases severely when the privacy requirements increase so that the model can't be applied directly to the case where only the pretrained models are given. \cite{papernot2016semi,papernot2018scalable} proposed a semi-supervised learning framework called PATE to reduce the impact of the DP noise by leveraging the noisy aggregation of multiple teacher models trained directly on the private data. It is possible to train a privacy-preserving student model using PATE framework given only sensitive teacher models, but it is difficult to find a suitable unlabeled public dataset for the distillation process.
    
    In the meantime, an independent line of research concerning model compression shows that some data-free knowledge distillation approaches~(DFKD)~\cite{chen2019data,zhu2021data,choi2020data} could achieve similar performance by vanilla training with only a teacher model. The data used for the distillation process is generated by a generator, which could potentially be a remedy for the above problem of the suitable public dataset. The generators of such methods mainly learn the data distribution rather than the image details, which intuitively also provides a degree of protection of privacy. \cite{Ge2023TIP} has showed that: {It is possible to leverage the power of data-free knowledge distillation to train a privacy-preserving student model that is not necessary to access to the original dataset.} But there is still a long way to take advantage of this intuition.
    
    Inspired by the above observations, in this paper we propose a model conversion approach with \textit{Differentially Private Data-Free Distillation} (DPDFD) to facilitate the model releasing by distilling a pretrained model as teacher into a differentially private student. Specifically, our DPDFD combines DFKD and DP, which applies DFKD to distill private knowledge and a DP mechanism $\mathcal{A}_{C,\sigma}$ to guarantee privacy. The objective is to enable an effective conversion that achieves strong privacy protection with minimum accuracy loss when only private models (teachers) are given. As shown in Fig.~\ref{Fig:framework}, we first generate massive synthetic data with a generator. Then, we feed the synthetic data into the teacher model and student model to compute the loss. Differentially private gradients are calculated by applying DP mechanism $\mathcal{A}_{C,\sigma}$. Finally, we update the student with these gradients and update the generator by taking the student as a fixed discriminator. In particular, we achieve DP by performing normalization on the gradients of student outputs and adding Gaussian noise to gradients during student learning. The reason for performing normalization instead of clipping is that it will retain the relative size information of the gradients and achieve better performance with a smaller norm bound. The reason for adding Gaussian noise to the gradients of student outputs is that it has a lower dimension compared to other gradients. Both smaller norm bound and lower dimension gradients make it easier to balance the performance and privacy. In addition, the DP mechanism $\mathcal{A}_{C,\sigma}$ also ensures differentially private training of the generator according to the post-processing mechanism. We can use the generator to generate data for other down stream tasks if needed. We also provide privacy and convergence analysis for our DPDFD in theory. Furthermore, DPDFD can be extended to multi-model case, which aggregates multiple sensitive teachers into a privacy-preserving student model.
    
    In summary, our DPDFD can effectively convert a sensitive teacher model to a privacy-preserving student model through three key components. First, performing normalization instead of clipping which is usually used in other approaches retains information about the relative size of the gradients. Second, achieving DP by adding noise on the gradients of lower dimensional outputs makes it easier to balance performance and privacy. Third, synthetic data generated by the generator has a similar distribution with the training data of teacher model. In this way, we can convert the sensitive model to a privacy-preserving student with minimum accuracy loss.
    
    Our major contributions are three folds: 1) we propose a model conversion approach to convert a pretrained  sensitive model into a privacy-preserving model for secure releasing; 2) we provide privacy analysis and convergence analysis for our approach in theory; 3) we conduct extensive experiments and analysis to demonstrate the scalability and effectiveness of our approach.

\section{Related Works}
    The approach we proposed in this paper aims to train privacy-preserving student models by distilling knowledge from given sensitive model(s) via a differentially private data-free distillation. Therefore, we briefly review the related works for two inspects, including {differentially private learning} and {data-free knowledge distillation}.
    
  \subsection{Differentially Private Learning}
    Differentially private learning aims to ensure the learning model is differentially private regarding the private data. \cite{abadi2016deep} proposed a differentially private stochastic gradient descent~(DPSGD) algorithm which achieved DP by clipping and adding noise to the gradients of all parameters during the training process. However, the model performance degrades severely with strong privacy requirements.  \cite{papernot2016semi} later proposed PATE which used semi-supervised learning to transfer the knowledge of the teacher ensemble to the student by using a noisy aggregation. It can reduce the impact of noise on performance by increasing the number of teacher models. However, it is difficult to find an unlabeled public dataset that has similar distribution to the training data of teachers. Some works want to train differentially private generators to generate data that has similar distribution to private data while preserving privacy. \cite{xie2018differentially} applied DPSGD to the training process of Generative Adversarial Networks~(GAN)~\cite{goodfellow2014generative} to get a differentially private generator. \cite{chen2020gs} suggested that it is not necessary to clip and add noise to all gradients, but only needs to achieve DP in the back propagation process from discriminator to generator. \cite{cao2021don} applied differentially private optimal transmission theory to train generators. \cite{chen2022dpgen} proposed an energy-guided network trained on sanitized data to indicate the direction of the true data distribution via Langevin Markov Chain Monte Carlo sampling method. In this paper, we make a better balance between performance and privacy by applying normalization instead of clipping and post-processing of DP.
    
    \begin{figure*}[t]
    \centering
    \includegraphics[width=1.0\linewidth]{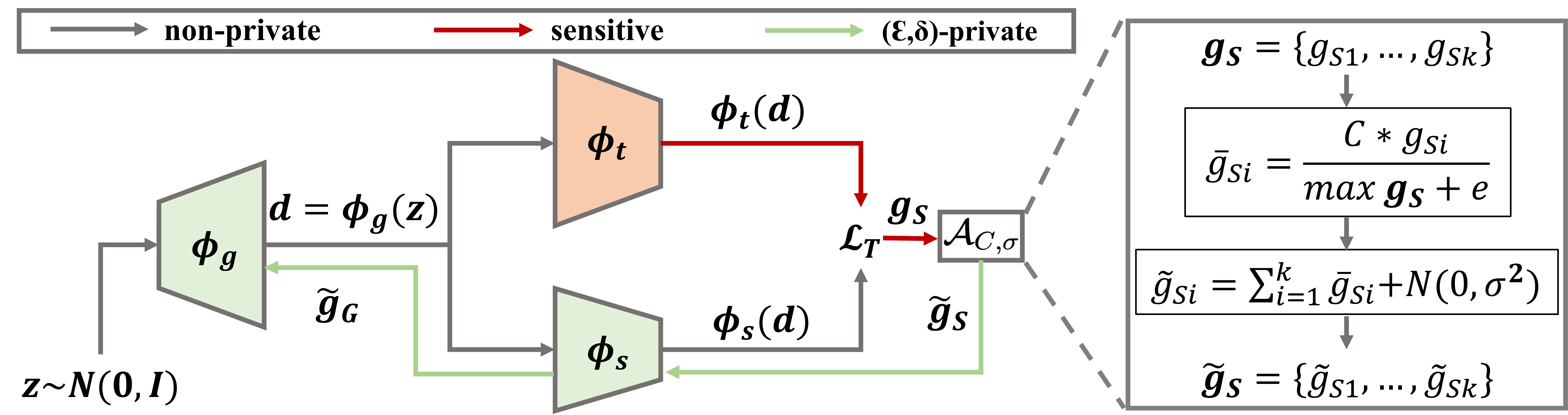}
    \caption{Overview of our differentially private data-free distillation approach. The approach learns to convert a pretrained model $\phi_t$ into a privacy-preserving student $\phi_s$ via an intermediate generator $\phi_g$. The learning is performed to collaborate three parties in a unified way. First, the generator generates massive data. Then, these data are fed into the teacher and student models to calculate the gradients $g_s$; Finally, the student and generator are updated with differentially private gradients $\tilde{g}_s$, which are computed by applying DP mechanism $\mathcal{A}_{C,\sigma}$ to $g_s$. Here, $C$ is the norm bound and $N(0,\sigma^2)$ is Gaussian noise with mean 0 and variance $\sigma^2$.
    }
    \label{Fig:framework}
    \end{figure*}
    
  \subsection{Data-Free Knowledge Distillation}
    Data-free knowledge distillation is a class of approaches that aims to train a student model with a pretrained teacher model without access to original training data. It uses the information extracted from the teacher model to synthesize data used in the distillation process. \cite{srinivas2015data} proposed to directly merge similar neurons in fully-connected layers, which cannot be applied to convolutional layers and networks because the detailed architectures and parameters information is unknown. \cite{lopes2017data} first proposed data-free knowledge distillation, using the distribution information of the original data to reconstruct the synthetic data used in the distillation process. \cite{deng2021graph} used this approach to generate synthetic graph data for data-free knowledge distillation for graph classification tasks. \cite{zhu2021data} modified and applied it to federated settings. \cite{chen2019data} proposed a novel framework named DAFL for training the student model by exploiting GAN. It uses the teacher model as the discriminator to train a generator. \cite{choi2020data} proposed matching statistics from the batch normalization layers for generated data and the original data in the teacher. It can make the generated data closer to the original data. \cite{fang2022up} proposed FastDFKD that applied the idea of meta-learning to the training process to accelerate the efficiency of data synthesis. Inspired by these approaches, it is possible to convert a sensitive model into a privacy-preserving model without access to the original data.

\section{Preliminaries}
    Here we will first provide some background knowledge about differential privacy. 
    We then draw connections between the definitions and theorems we introduced here and our DP analysis on DPDFD later. 
    The following definition explains how DP provides rigorous privacy guarantees clearly.
    \begin{definition}[Differential Privacy]\label{def:dp}
    \itshape{A randomized mechanism $\mathcal{A}$ with domain $\mathcal{R}$ is $(\varepsilon, \delta)$-differential privacy, if for all $\mathcal{O} \subseteq \mathcal{R}$ and any adjacent datasets $D$ and $D^{\prime}$ :}
    \begin{equation}
        \begin{aligned}
        Pr[\mathcal{A}(D) \in \mathcal{O}] \leq e^\varepsilon \cdot Pr\left[\mathcal{A}\left(D^{\prime}\right) \in \mathcal{O}\right]+\delta,
        \end{aligned}
    \end{equation}
    \end{definition} 
    \noindent where adjacent datasets $D$ and $D^{\prime}$ differ from each other with only one training example. $\varepsilon$ is the privacy budget and the smaller it is the better, and $\delta$ is the failure probability.

    \cite{mironov2017renyi} proposed a variant R$\acute{e}$nyi differential privacy~(RDP), which computes privacy with R$\acute{e}$nyi divergence. We review the definition of RDP and its connection to DP.
    \begin{definition}[R$\acute{\textbf{e}}$nyi Differential Privacy]\label{def:rdp}
    \itshape{A randomized mechanism $\mathcal{A}$ is $(\lambda, \varepsilon)$-RDP with $\lambda > 1$ if for any adjacent datasets $D$ and $D^{\prime}$ :}
    \begin{equation}
        \begin{aligned}
        D_{\lambda}&(\mathcal{A}(D)||\mathcal{A}(D^{\prime}))=\\&\frac{1}{\lambda-1}\log \mathbb{E}_{(x\sim\mathcal{A}(D))}\left[\left(\frac{Pr[\mathcal{A}(D)=x]}{Pr[\mathcal{A}(D^{\prime})=x]}\right)^{\lambda-1}\right]\le \varepsilon.
        \end{aligned}
    \end{equation}
    \end{definition}
    Different from DP, RDP has a more friendly composition theorem and can be applied to both data-independent and data-dependent settings. It supports a tighter composition of privacy budget which can be described as follows: For a sequence of mechanisms $\mathcal{A}_1,...,\mathcal{A}_i,...,\mathcal{A}_k$, where $\mathcal{A}_i$ is $(\lambda, \varepsilon_i)$-RDP, the composition of them $\mathcal{A}_1\circ...\mathcal{A}_i...\circ\mathcal{A}_k$ is $(\lambda,\sum_i\varepsilon_i)$-RDP. $\mathcal{A}_i$ represents one query to the teacher model in our case. Moreover, the connection between RDP and DP can be described as follows:
    \begin{theorem}[Convert RDP to DP]\label{th:rdp-dp}
    \itshape{A $(\lambda, \varepsilon)$-RDP mechanism $\mathcal{A}$ also satisfies $(\varepsilon+\log \frac{\lambda-1}{\lambda}-\frac{\log \delta + \log \lambda}{\lambda-1}, \delta)$-DP.}
    \end{theorem}

    To provide DP guarantees, we exploit the post-processing~\cite{dwork2014algorithmic} described as follows:
    \begin{theorem}[Post-processing]\label{th:post-processing}
    \itshape{If mechanism $\mathcal{A}$ satisfies $(\varepsilon, \delta)$-DP, the composition of a data-independent function $\mathcal{F}$ with $\mathcal{A}$ is also $(\varepsilon, \delta)$-DP.}
    \end{theorem}

\section{Proposed Approach}

  \subsection{Problem Formulation}
    Given a sensitive teacher model $\phi_t$ with parameters $\theta_t$, the objective is converting it into a privacy-preserving student model $\phi_s$ with parameters $\theta_s$ that does not reveal data privacy and has the ability to perform similarly to the teacher model. To achieve that, we introduce a differentially private data-free distillation approach. We first sample a batch of noise vectors $\mathbf{z}=\{z_i\}_{i=1}^B$ and feed them into the generator $\phi_g$ with parameters $\theta_g$ to generate massive synthetic data $D=\{d_i\}_{i=1}^{B}$. Enter them into the teacher model and student model to calculate the loss $\mathcal{L}_T(\phi_t(\theta_t;D), \phi_s(\theta_s;D))$ and then calculate updated outputs of student $y_s$ to achieve DP with a differentially private mechanism $\mathcal{A}_{C,\sigma}$. Finally update the student  by the loss $\mathcal{L}_{S}(\phi_s(\theta_s;D),y_s)$ . Thus, the converting process can be formulated by minimizing an energy function $\mathbb{E}$:
    \begin{equation}
        \begin{aligned}
        \mathbb{E}&(\theta_s;\theta_t)=\mathbb{E}(\phi_s(\theta_s;D), y_s)\\=\mathbb{E}&\left(\phi_s(\theta_s;D),\phi_s(\theta_s;D)-\gamma\cdot \mathcal{A}_{C,\sigma}(g)\right),
        \end{aligned}
    \end{equation}
    where $g=\frac{\partial \mathcal{L}_T(\phi_t(\theta_t;D), \phi_s(\theta_s;D))}{\partial \phi_s(\theta_s;D)}$ and $\gamma$ is the learning rate. We suppress the risk of privacy leakage with DP mechanism $\mathcal{A}_{C,\sigma}$ which normalizes the gradients with norm bound $C$ and adds Gaussian noise with variance $\sigma^2$. We will introduce how the differentially private mechanism $\mathcal{A}_{C,\sigma}$ protects privacy in detail in the following.
    
    \begin{algorithm}[tb]
    \caption{DPDFD}
    \label{alg:DPDFD}
    \textbf{Input}: Training iterations $T$, loss function $\mathcal{L}_T, \mathcal{L}_S, \mathcal{L}_G$, noise scale $\sigma$, sample size $B$, learning rate $\gamma, \gamma_s, \gamma_g$, gradient norm bound $C$, a positive stability constant $e$\\
    \begin{algorithmic}[1] 
    \FOR {$t \in [T]$}
    \STATE Sample $B$ noise samples $\textbf{z}=\{z_i\}_{i=1}^{B}$
    \STATE Generate $B$ synthetic samples $D=\{\phi_g(\theta_g;z_i)\}_{i=1}^{B}$
    \FOR {each synthetic data $d_i=\phi_g(\theta_g;z_i)$}
    \STATE Compute loss $\mathcal{L}_T(\phi_t(\theta_t;d_i),\phi_s(\theta_s;d_i))$
    \STATE Compute the gradient $g_i=\frac{\partial \mathcal{L}_T}{\partial \phi_s(\theta_s;d_i)}$
    \STATE Normalize the gradient $\bar{g}_i=\frac{C\cdot g_i}{||g||_2+e}$
    \ENDFOR
    \STATE Add noise $\tilde{g}=\frac{1}{B}(\sum\limits^{B}\limits_{i=1}\bar{g}_i + \mathcal{N}(0,\sigma^2C^2I))$
    \STATE Compute differentially private outputs of student $y_s=\phi_s(\theta_s;D)-\gamma\cdot \tilde{g}$
    \STATE Compute loss $\mathcal{L}_S(\phi_s(\theta_s;D), y_s)$
    \STATE Update student $\theta_s^{t+1}=\theta_s^{t}-\gamma_s\cdot \frac{\partial \mathcal{L}_S}{\partial \theta_s^{t}}$
    \STATE Compute loss $\mathcal{L}_G(\phi_s(\theta_s;D))$
    \STATE Update generator $\theta_g^{t+1}=\theta_g^{t}-\gamma_g \cdot\frac{\partial (\mathcal{L}_S+\mathcal{L}_G)}{\partial \theta_g^{t}}$
    \ENDFOR
    \STATE \textbf{return} $\theta_s$ and $\theta_g$
    \end{algorithmic}
    \end{algorithm}

  \subsection{Our Solution: DPDFD}
    Student model distilled from a sensitive teacher model directly may lead to privacy leakage and another main problem is that we don't have access to the original dataset. Thus, we aim to convert the sensitive teacher model to a privacy-preserving student model while having similar performance to the sensitive model in a data-free manner.
    
    Unlike \cite{abadi2016deep} which requires clipping and adding Gaussian noise to gradients of all parameters, due to the post-processing of DP, we only need to perform normalization and add noise to the gradients of student's outputs to calculate new differentially private outputs. As shown in Fig.~\ref{Fig:framework} and Alg.~\ref{alg:DPDFD}, we first sample a batch of noise vectors $\mathbf{z}=\{z_i\}_{i=1}^{B}$ and feed them into the generator $\phi_g$ with parameter $\theta_g$ to obtain massive synthetic data $D=\phi_g(\mathbf{z})$. Then enter the synthetic data $D$ into the teacher and student to compute the loss $\mathcal{L}_T(\phi_t(\theta_t;D),\phi_s(\theta_s;D))$. To get better results, we treat $\mathrm{argmax}(\phi_t(\theta_t;D))$ as the target labels and then calculate the distillation loss $\mathcal{L}_T$ same as \cite{zhao2022decoupled}. After that we achieve DP by the mechanism $\mathcal{A}_{C,\sigma}$ which can be described as follows:
    \begin{equation}\label{eq:dp}
        \begin{aligned}
        \mathcal{A}_{C,\sigma}(g)=\frac{1}{B}(\sum\limits^{B}\limits_{i=1}\frac{C\cdot g_i}{||g||_2+e} + \mathcal{N}(0,\sigma^2C^2I)),
        \end{aligned}
    \end{equation}
    and generate new differentially private outputs $y_s$. Compared with clipping, normalization can achieve higher accuracy at smaller norm bound $C$. This is because when $C$ is small, clipping makes the gradients lose their variability but normalization retains the relative size relationship of gradients. Another important point is that the smaller the $C$, the smaller the privacy budget consumed by each query to the teacher model, which is exactly what we want. So we choose a smaller $C$ and normalization operation to get a lower privacy budget and better performance. Finally,  we compute the loss $\mathcal{L}_S(\phi_s(\theta_s;D),y_s)$
    and update the student model with it. This loss function can take many forms, and we use cross-entropy loss in our experiments. In order to make the distribution of the synthetic data closer to the private data and balance the classes of the synthetic data, we add an additional loss $\mathcal{L}_G$ when updating the generator. It can be formulated as:
    \begin{equation}
        \begin{aligned}
        \mathcal{L}_G(\phi_s(\theta_s;D))=&\ell(\phi_s(\theta_s;D),\mathrm{argmax}(\phi_s(\theta_s;D)))\\+&\alpha\phi_s(\theta_s;D)\log(\phi_s(\theta_s;D))\\+&\beta||\phi_s(\theta_s^{-};D)||_2 ,
        \end{aligned}
    \end{equation}
    where $\alpha$ and $\beta$ are the tuning parameters to balance the effect of three terms, and we set both of them to 1. The first term $\ell(\cdot)$ is a cross entropy function that measures the one-hot classification loss, which enforces the synthetic data having similar distribution as the private data. The second term is the information entropy loss to measure the class balance of generated data. The third term uses $l_2$-norm $||*||_2$ to measure the activation loss, since the features that are extracted by the student and correspond to the output before the fully-connected layer tend to receive higher activation value if input data are real rather than some random vectors, where $\theta_s^{-} \subset \theta_s$ is the student’s backbone parameters. We update the student and generator alternately in our experiments. In the nutshell, the trained student and generator satisfy DP because the training process can be seen as post-processing, given that $y_s$ is differentially private results. Overall, DPDFD can be formally proved DP in Theorem.~\ref{th:dp-theorem}.

    \begin{theorem}\label{th:dp-theorem}
    \itshape{DPDFD guatantees $(\frac{2C^2nBT\lambda}{\sigma^2}+\log \frac{\lambda-1}{\lambda}-\frac{\log \delta + \log \lambda}{\lambda-1},\delta)$-DP for all $\lambda\ge 1$ and $\delta\in(0,1)$.}
    \end{theorem}
    

  \subsection{Convergence Analysis}
    To analyze the convergence of our DPDFD, we follow the standard assumptions in the SGD literature~\cite{allen20172,bottou2018optimization,ghadimi2013stochastic}, with an additional assumption on the gradient noise. We assume that $\mathcal{L}_T$ has a lower bound $\mathcal{L}_{*}$ and $\mathcal{L}_T$ is $\kappa$-smoothness, which can be described as follows: $\forall x,y$, there is an non-negative constant $\kappa$ such that $\mathcal{L}_T(x)-\mathcal{L}_T(y) \le \nabla \mathcal{L}_T(x)^{\top}(x-y) + \frac{\kappa}{2}||x-y||^2$. The additional assumption is that $(g_r-g)\sim \mathcal{N}(0,\zeta^2)$, where $g_r$ is the ideal gradients of $\mathcal{L}_T$ and $g$ is the gradient we compute as an unbiased estimate of $g_r$. Then according to~\cite{bu2022automatic}, in our case we have:
    \begin{equation}
    \min_{0 \leq t \leq T} \mathbb{E}\left(\left\|g_{r}^t\right\|\right) \leq \mathcal{F}\left( \sqrt{\frac{2\left(\mathcal{L}_{0}-\mathcal{L}_{*}\right)+2T\kappa \gamma^2C^2(1+\sigma^2d)}{T\gamma C}} ; \zeta, e\right),
    \end{equation}
    where $d$ is a constant number, $\mathcal{L}_0$ is the initial loss and $\mathcal{F}(\cdot)$ results only from the normalization operation same as~\cite{bu2022automatic} and it won't affect the monotonicity of input variables. We simply set the learning rate $\gamma \propto \frac{1}{\sqrt{T}}$ and the gradients will gradually tend to 0 as $T$ increases.

\begin{table*}[!htbp]
 	\setlength{\tabcolsep}{2.0pt}%
 	\renewcommand\arraystretch{1.3}
	\begin{center}
		\begin{threeparttable}
			\begin{tabular}{l|c|c|ccccccccc|c}
				\toprule
				\textbf{Dataset}& \textbf{Teacher} &$\varepsilon$ & \textbf{DP-GAN} & \textbf{PATE-GAN} & \textbf{G-PATE} & \textbf{GS-WGAN} & \textbf{DP-MERF} & \textbf{P3GM} & \textbf{DataLens} & \textbf{DPSH}  & \textbf{DPGEN} & \textbf{DPDFD}\cr
				\hline
				\multirow{2}{*}{\textbf{MNIST}} & \multirow{2}{*}{\textbf{0.9921}} & 1 & 0.4036 & 0.4168 & 0.5810 & 0.1432 & 0.6367 & 0.7369 & 0.7123 & N/A & 0.9046 & \textbf{0.9512} \cr
				& & 10 & 0.8011 & 0.6667 & 0.8092 & 0.8075 & 0.6738 & 0.7981 & 0.8066 & 0.8320 & 0.9357 & \textbf{0.9751}\cr
				\hline
				\multirow{2}{*}{\textbf{FMNIST}} & \multirow{2}{*}{\textbf{0.9102}} & 1 & 0.1053 & 0.4222 & 0.5567 & 0.1661 & 0.5862 & 0.7223 & 0.6478 & N/A & 0.8283 & \textbf{0.8386}\cr
				& & 10 & 0.6098 & 0.6218 & 0.6934 & 0.6579 & 0.6162 & 0.7480 & 0.7061 & 0.7110 & 0.8784 & \textbf{0.8988}\cr
				\hline
				\multirow{2}{*}{\textbf{CelebA-G}} & \multirow{2}{*}{\textbf{0.9353}} & 1 & 0.5330 & 0.6068 & 0.6702 & 0.5901 & 0.5936 & 0.5673 & 0.7058 & N/A & 0.6999 & \textbf{0.7237}\cr
				& & 10 & 0.5211 & 0.6535 & 0.6897 & 0.6136 & 0.6082 & 0.5884 & 0.7287 & 0.7630 & 0.8835 & \textbf{0.8992}\cr
				\hline
				\multirow{2}{*}{\textbf{CelebA-H}} & \multirow{2}{*}{\textbf{0.8868}} & 1 & 0.3447 & 0.3789 & 0.4985 & 0.4203 & 0.4413 & 0.4532 & 0.6061 & N/A & 0.6614 & \textbf{0.7839}\cr
				& & 10 & 0.3920 & 0.3900 & 0.6217 & 0.5225 & 0.4489 & 0.4858 & 0.6224 & N/A & 0.8147 & \textbf{0.8235}\cr
				\bottomrule
			\end{tabular}
		\end{threeparttable}
	\end{center}
 \caption{Accuracy comparisons with 9 explicit approaches under different privacy budget $\varepsilon$ ($\delta=10^{-5}$).}\label{tab:dp(1_and_10)}
\end{table*}

  \subsection{Discussion}
    \myPara{Extension to Multi-Model Ensemble.}~Our DPDFD can be extended to the case of multiple sensitive models. Different from the case of a single teacher that reduces the impact of DP noise by averaging a batch of gradients, the case of multi-model achieves it by averaging the gradients from given sensitive models. In particular, given multiple teacher models $\{\phi_t^j\}_{j=1}^{n}$, we first sample noise vectors and generate massive synthetic data same as the case of a single teacher. For each data $d_i$, we enter it into multiple teacher models $\{\phi_t^j(\theta_t^j;d_i)\}_{j=1}^{n}$ and student model $\phi_s(\theta_s;d_i)$ and compute losses $\{\mathcal{L}_T(\phi_t^j(\theta_t^j;d_i),\phi_s(\theta_s;d_i))\}_{j=1}^{n}$. Then, we compute gradients $g_{ij}=\left\{\frac{\partial \mathcal{L}_T(\phi_t^j(\theta_t^j;d_i),\phi_s(\theta_s;d_i))}{\partial \phi_s(\theta_s;d_i)}\right\}_{j=1}^{n}$ with the losses. After that, we normalize them with norm bound $C$ and add noise to them to get differentially private gradients $\tilde{g}_i=\frac{1}{n}(\sum\limits^{n}\limits_{i=1}\frac{C\cdot g_{ij}}{||g||_2+e} + \mathcal{N}(0,\sigma^2C^2I))$. Next we can compute an updated output of student $y_s^i=\phi_s(\theta_s;d_i)-\gamma \cdot\tilde{g}_i$. Finally, we update the student and generator same as Alg.~\ref{alg:DPDFD}. In this way, we can aggregate multiple sensitive teacher models into a privacy-preserving student model. 

    \myPara{Direct Training on Private Data.}~The proposed DP mechanism $\mathcal{A}_{C,\sigma}$ is a key building block of DPDFD, and can also be applied to situations where private data is available. Given a private dataset $\{\mathbf{x},y\}$, we input the data $\mathbf{x}$ into the model and calculate the cross-entroy loss $\mathcal{L}( \phi(\theta;\mathbf{x}),y)$. After that, we run the same procedures to achieve DP and update the model as Alg.~\ref{alg:DPDFD} except we don't need to update the generator. In this way, we can train a privacy-preserving model with direct access to private data.

\section{Experiments}
    In this section, we present the experimental evaluation of DPDFD for converting the sensitive model to a differentially private model with high performance.

  \subsection{Experimental Setup}
    \myPara{Datasets.}~We conduct our experiments on 7 image datasets, including MNIST~\cite{lecun1998gradient}, FashionMNIST(FMNIST)~\cite{fashionMnist}, CIFAR10~\cite{cifar10}, CelebA~\cite{liu2015celeba}, PathMNIST~\cite{medmnistv2}, COVIDx and ImageNet~\cite{deng2009imagenet}. We created CelebA-H and CelebA-G based on CelebA. CelebA-H and CelebA-G are classification datasets with hair color~(black/blonde/brown) and gender as the label, respectively. COVIDx is a classification dataset for COVID.
    
    \myPara{Baselines.}~We perform compartments with 14 state-of-the-art benchmarks, including 9 explicit approaches that training classifiers with generative data~(DP-GAN~\cite{xie2018differentially}, GS-WGAN~\cite{chen2020gs}, PATE-GAN~\cite{jordon2018pate}, DP-MERF~\cite{harder2020differentially}, P3GM~\cite{takagi2021p3gm}, DataLens~\cite{wang2021ccs}, G-PATE~\cite{long2021g}, DPSH~\cite{cao2021don}, DPGEN~\cite{chen2022dpgen}) and 5 implicit approaches that training classifiers with differentially private learning~(DPSGD~\cite{abadi2016deep}, TSADP~\cite{papernot2021tempered}, TOPAGG~\cite{wang2021ccs}, GM-DP~\cite{mcmahan2018general}, DGD~\cite{Ge2023TIP}). To make the comparisons fair, we take the results from original papers or run the official codes. 

    \myPara{Networks.}~We adopt several popular network architectures as our teacher models, including AlexNet~\cite{krizhevsky2012imagenet}, VGGNet~\cite{vgg2015iclr}, ResNet~\cite{resnet2016cvpr}, WideResnet~\cite{zagoruyko2016wide}, DenseNet~\cite{huang2016deep}, MobileNet~\cite{howard2017mobilenets}, ShuffleNet~\cite{zhang2018shufflenet}, GoogleNet~\cite{szegedy2015going} and ViT~\cite{dosovitskiy2020image}. For VGGNet, we use 19-layer net with BN. For ResNet, we use 50-layer networks for ImageNet and 34-layer networks for others. For WideResnet, we use 50-layer networks as teacher. For DenseNet, we use 161-layer networks with growth rate equals 24. For ViT, we use the same architecture as~\cite{dosovitskiy2020image}. For student model, we use 34-layer ResNet for ImageNet and the same network as DataLens for others.

  \subsection{Experimental Results}
    In this section, we compare DPDFD with 14 baselines and evaluate it on different networks to verify its effectiveness. We first compare the model performance of DPDFD and other state-of-the-art approaches. Then, we conduct experiments on ImageNet with different networks under different privacy budget. We show that our DPDFD is scalable and outperforms all baselines. 

    \myPara{Comparisons with 9 Explicit Baselines.}~To demonstrate the effectiveness of our DPDFD, we conduct comparisons with 9 baselines under different privacy budget and report the results in Tab.~\ref{tab:dp(1_and_10)}. All approaches are under a low failure probability $\delta=10^{-5}$. We can see that our DPDFD achieves the highest performance under the same condition of privacy budget. In particular, when $\varepsilon=1$, our DPDFD achieves an accuracy of 0.9512 on MNIST and 0.8386 on FMNIST, which remarkably reduces the accuracy drop by 0.0409 and 0.0716 respectively, while most of the other approaches fail to achieve an accuracy of 0.8000. It shows that our approach has the best privacy-preserving ability and minimal accuracy drop. Even for high dimensional datasets like CelebA whose dimensionality is about 16 times larger than MNIST, all approaches suffer from accuracy drop with respect to their counterparts under high privacy budget while our DPDFD still delivers the highest test accuracy. This demonstrates that our approach is also effective for high-dimensional datasets. 

    \myPara{Comparisons with 5 Implicit Baselines.}~In addition, we conduct experimental comparisons with 5 implicit baselines on MNIST and CIFAR10. The results are shown in Tab.~\ref{tab:implicit}, where our DPDFD achieves the highest accuracy of 0.9512 and 0.8601 under the lowest privacy budget of 1.0 and 2.0. The main reason comes from that we use normalization instead of traditional clipping. In this way, the differentially private gradients obtained after $\mathcal{A}_{C,\sigma}$ retain more information~(relative size between gradients).

\begin{table}[!htbp]
 	\setlength{\tabcolsep}{3.0pt}%
 	\renewcommand\arraystretch{1.35}
	\begin{center}
		\begin{threeparttable}
			\begin{tabular}{l|c|c|c|c}
				\toprule
				\textbf{Approach} & $\varepsilon$ & \textbf{MNIST} & $\varepsilon$ & \textbf{CIFAR10}\cr
				\hline
				\textbf{DPSGD} & 2.0 & 0.9500 & 2.0 & 0.6623\cr
                    \textbf{TSADP} & 1.0 & 0.7991 & 7.5 & 0.6620\cr
                    \textbf{TOPAGG}& 1.0 & 0.9465 & 2.0 & 0.8518\cr
                    \textbf{GM-DP} & 1.0 & 0.9508 & 2.0 & 0.8597\cr
                    \textbf{DGD}  & 1.0 & 0.7360 & 3.0 & 0.7365\cr
                    \hline
                    \textbf{DPDFD}& 1.0 & \textbf{0.9512} & 2.0 & \textbf{0.8601}\cr
				\bottomrule
			\end{tabular}
		\end{threeparttable}
	\end{center}
 \caption{Accuracy comparisons with 5 implicit approaches under different privacy budget $\varepsilon$.}\label{tab:implicit}
\end{table}

\begin{table}[!htbp]
 	\setlength{\tabcolsep}{3.0pt}%
 	\renewcommand\arraystretch{1.35}
	\begin{center}
		\begin{threeparttable}
			\begin{tabular}{l|c|ccc}
				\toprule
				\multirow{2}{*}{\textbf{Networks}} & \multirow{2}{*}{\textbf{Teacher}} & \multicolumn{3}{c}{$\varepsilon$} \cr 
				\cline{3-5}
				 &  & 1 & 2 & 5\cr
				\hline
				\textbf{AlexNet} & \textbf{0.5655} & 0.2756 & 0.5124 & 0.5218\cr
				\hline
				\textbf{VGGNet} & \textbf{0.7423} & 0.3419 & 0.6381 & 0.6602\cr
				\hline
				\textbf{ResNet} & \textbf{0.7602} & 0.3823 & 0.6499 & 0.6781\cr
				\hline
				\textbf{WideResNet} & \textbf{0.7848} & 0.2982 & 0.5997 & 0.7117\cr
				\hline
				\textbf{DenseNet} & \textbf{0.7765} & 0.4552 & 0.6612 & 0.7029\cr
				\hline
				\textbf{MobileNet} & \textbf{0.7186} & 0.3830 & 0.6203 & 0.6424\cr
				\hline
				\textbf{ShuffleNet} & \textbf{0.6953} & 0.4964 & 0.6313 & 0.6652\cr
				\hline
				\textbf{GoogleNet} & \textbf{0.6246} & 0.2987 & 0.5717 & 0.5858\cr
				\hline
				\textbf{ViT} & \textbf{0.8140} & 0.3142 & 0.6880 & 0.7240\cr
				\bottomrule
			\end{tabular}
		\end{threeparttable}
	\end{center}
 \caption{Accuracy on ImageNet with different networks under different privacy budget $\varepsilon$~($\delta=10^{-5}$).}\label{tab:networks}
\end{table}

    \myPara{Evaluation on Open Source Networks.}~To demonstrate the scalability of our DPDFD, we apply it to convert several popular networks pretrained on ImageNet under different privacy budget $\varepsilon$. These teacher models are taken directly from the PyTorch model zoo. The results are shown in Tab.~\ref{tab:networks}. We find that the effect of network architectures is sometimes greater than the effect of public teacher model accuracy when $\varepsilon$ is small. In particular, when $\varepsilon=1$, the student model reaches the highest accuracy of 0.4964 when the teacher model is ShuffleNet, but the accuracy of teacher model is 0.1187 lower than the highest ViT. We guess this is because the simpler the teacher model is, the easier it is for the generator to learn the distribution of its training data, resulting in faster convergence of the student model. As $\varepsilon$ increases, the student model can learn more and more fully from the teacher model, so the effect of teacher performance dominates. The more complex the teacher model, the more information about the training data it contains. The student model learns more accurate predictions from the teacher and the generator learns a more similar data distribution, both of which lead to higher accuracy of the student model. We can see that at $\varepsilon=5$, the accuracy of the student is positively correlated with the accuracy of teacher model and it has approached that of the teachers, which proves that our DPDFD is not only scalable but also effective to a variety of popular network architectures.

\begin{table}[!htbp]
 	\setlength{\tabcolsep}{2.0pt}%
 	\renewcommand\arraystretch{1.35}
	\begin{center}
		\begin{threeparttable}
			\begin{tabular}{l|c|c|c|c|c}
				\toprule
				\textbf{Dataset} & $\varepsilon$ & \textbf{DPSGD} & \textbf{TOPAGG} & \textbf{GM-DP} & \textbf{DPDFD}\cr
				\hline
				\multirow{3}{*}{\textbf{MNIST}} & 0.5 & 0.8103 & 0.9235 & 0.9331 & \textbf{0.9377} \cr
				& 0.7 & 0.8932 & 0.9382 & 0.9438 & \textbf{0.9447} \cr
				& 1.0 & 0.9247 & 0.9465 & 0.9508 & \textbf{0.9762} \cr
				\hline
				\multirow{3}{*}{\textbf{CIFAR10}} & 2.0 & 0.6623 & 0.8518 & 0.8597 & \textbf{0.8652} \cr
				& 4.0 & 0.6884 & 0.8540 & 0.8663 & \textbf{0.8708} \cr
				& 8.0 & 0.7159 & 0.8562 & 0.8705 & \textbf{0.8973} \cr
				\bottomrule
			\end{tabular}
		\end{threeparttable}
	\end{center}
 \caption{Accuracy comparisons with 3 DPSGD mechanisms on MNIST and CIFAR10 under different privacy budget $\varepsilon$.}\label{tab:TDPD}
\end{table}

\myPara{Evalution of Training Directly with Private Data.}~In this section, we conduct experiments to evaluate the application of our DP mechanism $\mathcal{A}_{C,\sigma}$ to the case where private data can be accessed. We compare 3 state-of-the-art approaches which trained directly with private data~(based on DPSGD mechnism): DPSGD, TOPAGG and GM-DP on MNIST and CIFAR10 under the same settings except for the DP mechanism. The results are shown in Tab.~\ref{tab:TDPD}. We can see that our approach achieves the best performance on both the low-dimensional MNIST dataset and the high-dimensional CIFAR10 dataset. These results depend on two aspects. On one hand, we only need to compute a new differentially private output in the first step of backpropagation instead of adding noise to each gradient like other approaches. On the other hand, we use normalization instead of clipping, which retains more gradient information in the case of small norm bound $C$. So our approach can achieve better results.

  \subsection{Ablation Studies}
    After the promising performance is achieved, we further analyze the impact of each component in our approach, including normalization vs. clipping, norm bound, noise scale, and composition of loss function.

    \begin{figure}[t]
    \centering
    \includegraphics[width=1\columnwidth]{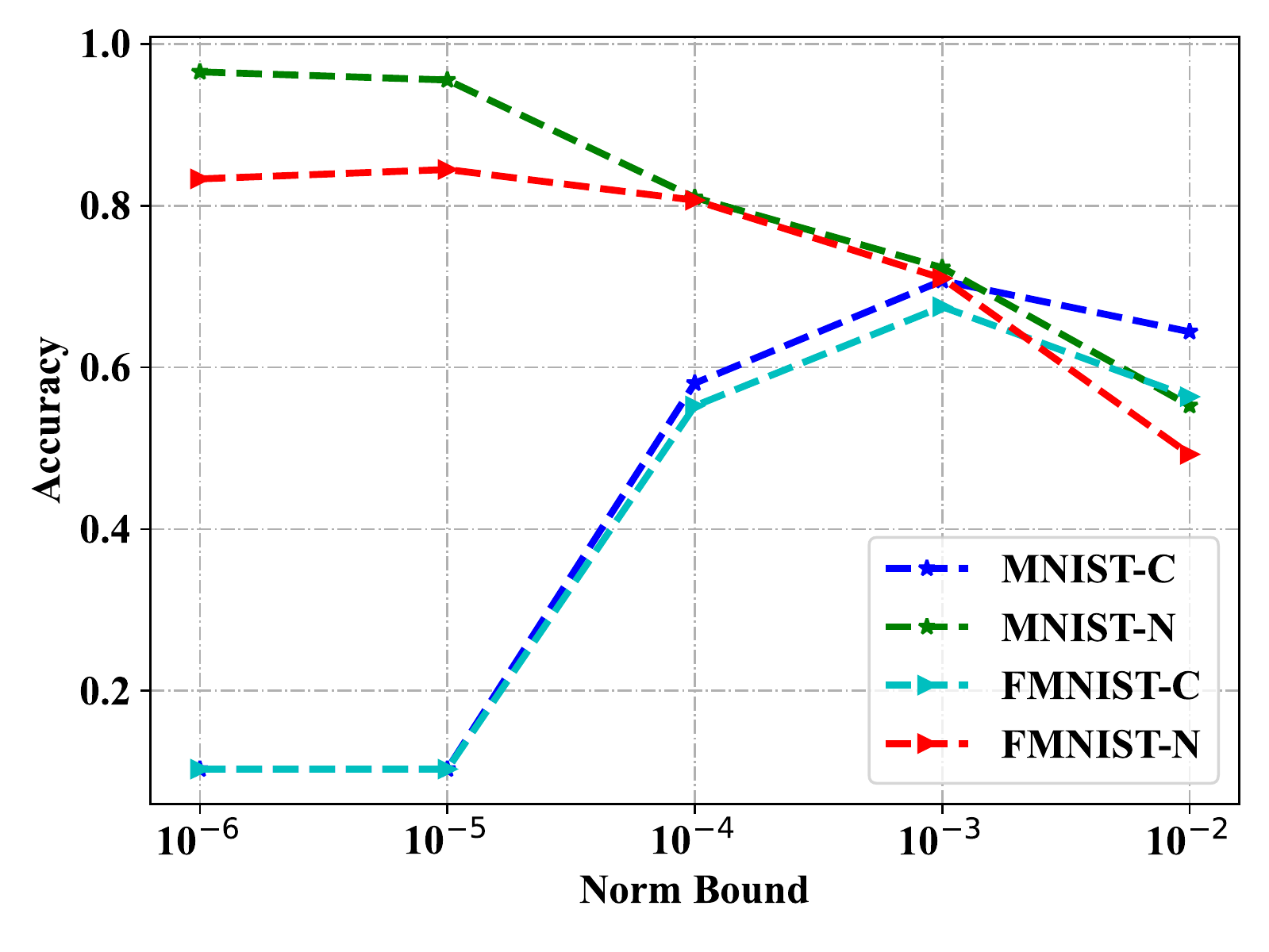}
    \caption{Student accuracy under different norm bound $C$ and different operations on gradients (`-C' or `-N') in the case $\varepsilon=0.6$ and $\delta=10^{-5}$ (`-C': Clipping, `-N': Normalization)}
    \label{Fig:NormBound}
    \end{figure}

    \myPara{Normalization vs. Clipping.}~To study the effect of different operations on the gradients, we conduct experiments with 34-layer ResNet pretrained on MNIST and FMNIST as teachers and report the results in Fig.~\ref{Fig:NormBound}. Performing normalization is significantly better than clipping when $C$ is small, and even clipping makes the student not converge. The advantage of normalization gradually decreases as $C$ increases, and when it increases to a certain level~(about $10^{-3}$ as Fig.~\ref{Fig:NormBound} shows), clipping will be superior to normalization. This is because normalization retains the relative size information of the gradients while clipping retains the absolute size information. 

    \myPara{Norm Bound.}~Norm bound is an important hyper-parameter of our DP mechanism $\mathcal{A}_{C,\sigma}$. We conduct experiments to investigate how norm bound $C$ affects the performance of student model. The results are shown as the red and green lines in Fig.~\ref{Fig:NormBound}. As we see, the student performs better when $C$ is small. This is because normalization preserves information about the relative size of the gradients. Although smaller $C$ will be more affected by DP noise, it will allow more training epochs. So the student performs better when $C$ is small.

    \begin{figure}[t]
    \centering
    \includegraphics[width=1\columnwidth]{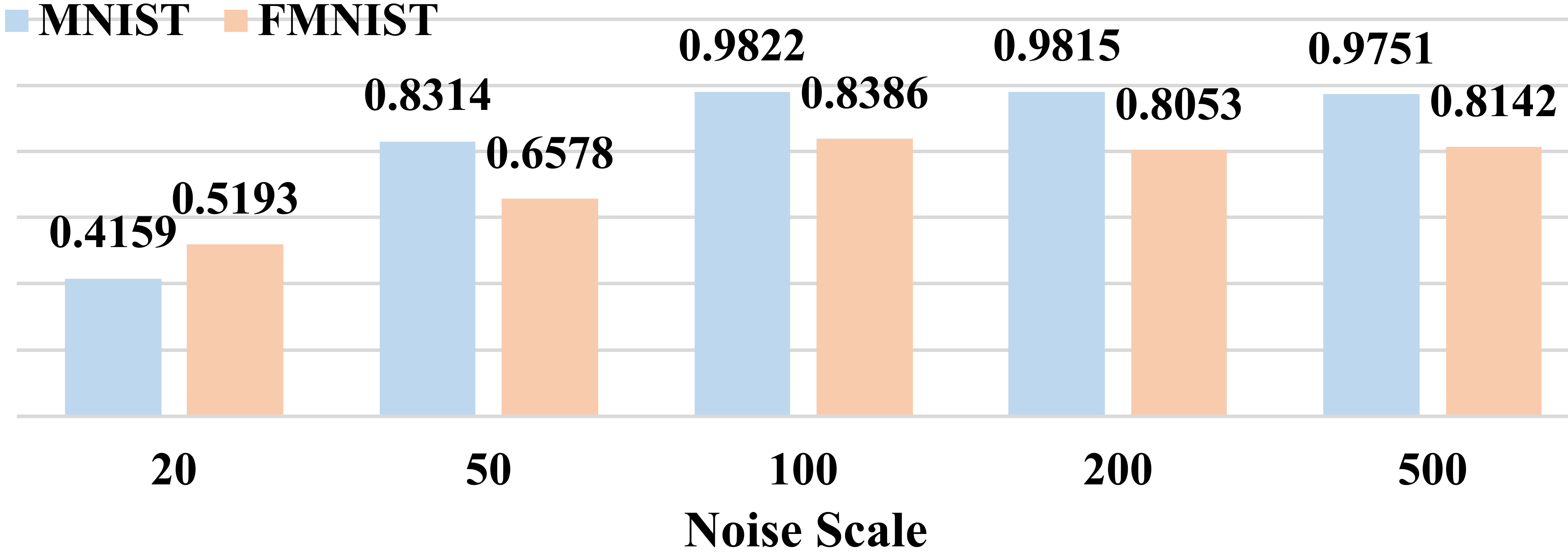}
    \caption{Student accuracy under different noise scale $\sigma$}
    \label{Fig:NoiseScale}
    \end{figure}

    \myPara{Noise Scale.}~Like norm bound, noise scale is also an important hyper-parameter of our DP mechanism $\mathcal{A}_{C,\sigma}$. To study the effect of it, we conduct experiments and the results are shown in Fig.~\ref{Fig:NoiseScale}. We observe that the model performance gets better as noise scale $\sigma$ increases from 20 to 100. This is because a larger noise scale consumes a smaller privacy budget per epoch, which allows the student model to learn more fully with a limited privacy budget. However, we find that there is a slightly worse decrease in model performance as noise scale increases from 100 to 500. This is because the gradients will be more broken with a larger noise scale, thus leading to a slightly worse performance. In practical applications, a trade-off should be made based on the actual situation.

\begin{table}[!htbp]
 	\setlength{\tabcolsep}{7.6pt}%
 	\renewcommand\arraystretch{1.35}
	\begin{center}
		\begin{threeparttable}
			\begin{tabular}{l|ccc|c}
				\toprule
				\textbf{Dataset}& \textbf{CE} & \textbf{IE} & \textbf{Norm} & \textbf{Accuracy} \cr
				\hline
				\multirow{4}{*}{\textbf{MNIST}} & \usym{2713} & \usym{2713} & \usym{2713} & \textbf{0.9751} \cr
				\cline{2-5}
				& \usym{2717} & \usym{2713} & \usym{2713} & 0.9691 \cr
				& \usym{2713} & \usym{2717} & \usym{2713} & 0.5463 \cr
				& \usym{2713} & \usym{2713} & \usym{2717} & 0.9496 \cr
				\hline
				\multirow{4}{*}{\textbf{FMNIST}} & \usym{2713} & \usym{2713} & \usym{2713} & \textbf{0.8988} \cr
				\cline{2-5}
				& \usym{2717} & \usym{2713} & \usym{2713} & 0.7649 \cr
				& \usym{2713} & \usym{2717} & \usym{2713} & 0.5463 \cr
				& \usym{2713} & \usym{2713} & \usym{2717} & 0.8806 \cr
				\bottomrule
			\end{tabular}
		\end{threeparttable}
	\end{center}
 \caption{Impact of each term in $\mathcal{L}_G$. The test accuracy of student models trained on synthetic data under $\varepsilon$=10 is reported. CE: Cross Entropy; IE: Information Entropy; Norm: $l_2$-Normalization.}\label{tab:loss}
\end{table}
    \myPara{Composition of Loss Functions.}~To check the effect of loss terms in the training generator, we investigate how each component of loss function contributes to the performance of student with 34-layer ResNet as the teacher model. We evaluate how they impact the performance by adding or removing each component and the results are shown in Tab.~\ref{tab:loss}. We note that the information entropy loss term is the most important component based on the results. This is because removing it will result in an imbalance in the classes of the synthetic data
    Cross entropy loss contributes differently for different datasets, with 1\% improvement on the MNIST dataset and 13\% improvement on the FMNIST dataset. $l_2$- normalization is also a useful term though less critical, contributing to the 2\%-3\% of the performance improvement as shown in Tab.~\ref{tab:loss}. 
    In summary, all three compositions have a positive effect on the performance of the converted model.

\section{Conclusion}
    Public pretrained models in model zoos may pose the risk of privacy leakage. To facilitate model deployment, we proposed a differentially private data-free distillation approach to convert sensitive teacher models into privacy-preserving student models. We train a generator for approximating the private dataset without the training data and student networks can be learned effectively through the knowledge distillation scheme. In addition, we perform normalization on the gradients of student outputs and add Gaussian noise to them to guarantee privacy. We also provide privacy analysis and convergence analysis for DPDFD. Extensive experiments are conducted to show the effectiveness of our approach. 
    In the future, we will explore the approach in more real-world applications like federated learning on medical images.

\myPara{Acknowledgements} 
    This work was partially supported by grants from the National Key Research and Development Plan (2020AAA0140001), and the Beijing Natural Science Foundation (19L2040).

\bibliographystyle{named}
\bibliography{dpdfd}
\newpage
\section*{Appendix}
\subsection*{Proof of Theorem. 3}
    We analyze the differential privacy bound for our proposed DPDFD in this section. Our approach is built on Gaussian mechanism~\cite{dwork2014algorithmic,mironov2017renyi} described as follows:
    \begin{theorem}[Gaussian Mechanism]\label{th:gaussian}
    \itshape{Let $f$ be a function with sensitive being $S_f=\max\limits_{D,D^{\prime}}||f(D)-f(D^{\prime})||_2$ over all adjacent datasets $D$ and $D^{\prime}$. The Gaussian mechanism $\mathcal{A}$ with adding noise to the output of $f$:$\mathcal{A}(x) = f(x) + \mathcal{N}(0, \sigma^2)$
    is $(\lambda, \frac{\lambda S_f^{2}}{2\sigma^2})$-RDP.}
    \end{theorem}
    We achieve the Gaussian mechanism by Eq.~\ref{eq:dp}.
    
    \begin{lemma}\label{lemma:sensitive}
    \itshape{For any neighboring gradient vectors $\mathcal{G}, \mathcal{G}^{\prime}$ differing by the gradient vector of one data with length $n$, the $l_2$ sensitivity is $2C\sqrt{n}$ after performing normalization with norm bound $C$.}
    \begin{proof}
    \itshape{The $l_2$ sensitivity is the max change in $l_2$ norm caused by the input change. For the vectors after normalization with norm bound $C$, each dimension has a maximum value of $C$ and a minimum value of $-C$. In the worst case, the difference of one data makes the gradient of all dimensions change from the maximum value $C$ to the minimum value $-C$, the change in $l_2$ norm equals $\sqrt{(2C)^2n}=2C\sqrt{n}$.}
    \end{proof}
    \end{lemma}

    \begin{theorem}\label{th:dp-proof}
    \itshape{DPDFD guatantees $(\frac{2C^2nBT\lambda}{\sigma^2}+\log \frac{\lambda-1}{\lambda}-\frac{\log \delta + \log \lambda}{\lambda-1},\delta)$-DP for all $\lambda\ge 1$ and $\delta\in(0,1)$.}
    \begin{proof}
    For each data, the gradient normalization and noise addition implements a Gaussian mechanism which guarantees $(\lambda, \frac{2C^2n\lambda}{\sigma^2})$-RDP (Theorem~\ref{th:gaussian} \& Lemma~\ref{lemma:sensitive}). So the DPDFD satisfies $(\lambda, \frac{2C^2nBT\lambda}{\sigma^2})$-RDP~(Theorem~\ref{th:post-processing} \& Composition of RDP), which is $(\frac{2C^2nBT\lambda}{\sigma^2}+\log \frac{\lambda-1}{\lambda}-\frac{\log \delta + \log \lambda}{\lambda-1},\delta)$-DP~(Theorem~\ref{th:rdp-dp}).
    \end{proof}
    \end{theorem}

\subsection*{More Details of Experimental Settings}
\myPara{Datasets.}~MNIST and FMNIST are both 10-class datasets containing $60K$ training examples and $10K$ testing examples. Each example is $28\times 28$ grayscale image. PathMNIST is a 9-class dataset for colon pathology, containing 89996 train data, 10004 valid data and 7180 test data, each sample is a $28\times28$ color images. CIFAR10 consists of $60K$ $32\times32$ color images in 10 classes, including $50K$ for training and $10K$ for testing. The COVIDx\footnote{https://www.kaggle.com/datasets/tawsifurrahman/covid19-radiography-database} dataset contains 19700 $299\times299$ color images and we split the train set and test set in 4:1. The CelebA dataset contains 202,599 color images of celebrity faces. We use the official preprocessed version with face alignment to resize the images to $64\times64\times3$. We create two CelebA datasets based on different attributes: CelebA-H and CelebA-G. CelebA-H uses three hair color attributes~(black/blonde/brown) as classification labels and CelebA-G uses male and female as classification labels. We partition them into train set and test set according to the official criteria~\cite{liu2015celeba}. 
ImageNet is a 1000-class dataset, containing about $135W$ high resolution color images.

\myPara{Implementation.}~For all datasets except ImageNet, we set the norm bound $C$ to $10^{-3}$ when $\varepsilon \le 1$ and $5\times10^{-3}$ when $1 < \varepsilon \le 10$. For ImageNet, we set the norm bound $C$ is $10^{-4}$. For all datasets, we set the noise scale $\sigma$, sample size $B$, positive stability constant $e$ and number of teachers $n$ to 100, 256, $10^{-4}$ and 100 correspondingly.

\subsection*{Applications to Medical Datasets}
We also conduct experiments with models pretrained on two medical datasets of real scenarios PathMNIST and COVIDx as teachers and compare with DataLens. The results are shown in Tab.~\ref{tab:dp(path_and_covidx)}. We can see that the performance of DPDFD on these two medical datasets is significantly better than that of the DataLens. These results imply that the important role of performing normalization instead of clipping and achieving DP for low-dimensional classifier outputs. The former can obtain more gradient information when norm bound $C$ is small, and the latter can get a smaller privacy budget when adding the same scale of noise.

\begin{table}[!htbp]
 	\setlength{\tabcolsep}{2.0pt}%
 	\renewcommand\arraystretch{1.25}
	\begin{center}
		\begin{threeparttable}
			\begin{tabular}{l|c|c|c|c}
				\toprule
				\textbf{Dataset} & \textbf{Teacher} & $\varepsilon$ & \textbf{DataLens} & \textbf{DPDFD}\cr
				\midrule
				\multirow{2}{*}{\textbf{PathMNIST}} & \multirow{2}{*}{\textbf{0.9009}} & 1 & 0.5352 & \textbf{0.8012} \cr
				& & 10 & 0.6890 & \textbf{0.8589}\cr
				\midrule
				\multirow{2}{*}{\textbf{COVIDx}} & \multirow{2}{*}{\textbf{0.8818}} & 1 & 0.4861 & \textbf{0.7361} \cr
				& & 10 & 0.5587 & \textbf{0.8169}\cr
				\bottomrule
			\end{tabular}
		\end{threeparttable}
	\end{center}
 \caption{Accuracy comparisons with DataLens on two medical datasets: Test accuracy under different privacy budget $\varepsilon$.}\label{tab:dp(path_and_covidx)}
\end{table}

\subsection*{Evaluation on Multi-Model Case}
We test the multi-model version of DPDFD and show the results in Tab.~\ref{tab:multi-model}. For fairness, the amount of data used for multi-model case is the same as single model case. Dividing a given dataset into disjoint n copies to train n teachers separately.
\begin{table}[!htbp]
 	\setlength{\tabcolsep}{2.0pt}%
 	\renewcommand\arraystretch{1.25}
	\begin{center}
		\begin{threeparttable}
			\begin{tabular}{l|c|c|c|c}
				\toprule
				\textbf{Dataset}& \textbf{Teacher} &$\varepsilon$ & \textbf{DPDFD} & \textbf{DPDFD-MM}\cr
				\midrule
				\multirow{2}{*}{\textbf{MNIST}} & \multirow{2}{*}{\textbf{0.9921}} & 1 & 0.9512 & 0.9321 \cr
				& & 10 & 0.9751 & 0.9556\cr
				\midrule
				\multirow{2}{*}{\textbf{FMNIST}} & \multirow{2}{*}{\textbf{0.9102}} & 1 & 0.8386 & 0.7993\cr
				& & 10 & 0.8988 & 0.8750\cr
				\midrule
				\multirow{2}{*}{\textbf{CelebA-G}} & \multirow{2}{*}{\textbf{0.9353}} & 1 & 0.7237 & 0.7312\cr
				& & 10 & 0.8992 & 0.8651\cr
				\midrule
				\multirow{2}{*}{\textbf{CelebA-H}} & \multirow{2}{*}{\textbf{0.8868}} & 1 & 0.7839 & 0.7781\cr
				& & 10 & 0.8235 & 0.8062\cr
				\bottomrule
			\end{tabular}
		\end{threeparttable}
	\end{center}
 \caption{Evaluation on multi-model case: Test accuracy under different privacy budget $\varepsilon$ ($\delta=10^{-5}$, DPDFD-MM: Multi-Model DPDFD).}\label{tab:multi-model}
\end{table}

\subsection*{Algorithms Details}
\myPara{Multi-Model DPDFD.}~We present the pseudocode of multi-model DPDFD in Alg.~\ref{alg:multi-df-dpd}.

\myPara{Training Directly with Private Data.}~We present the pseudocode of training directly with private data in Alg.~\ref{alg:TDPD}.

\begin{algorithm}[!htbp]
\caption{Multi-Model DPDFD}
\label{alg:multi-df-dpd}
\textbf{Input}: Number of training iterations $T$, loss function $\mathcal{L}_T, \mathcal{L}_S, \mathcal{L}_G$, noise scale $\sigma$, sample size $B$, learning rate $\gamma, \gamma_s, \gamma_g$, gradient norm bound $C$, a positive stability constant $e$, number of teachers $n$\\
\begin{algorithmic}[1] 
\FOR {$t \in [T]$}
\STATE Sample $B$ noise samples $\textbf{z}=\{z_i\}_{i=1}^{B}$
\STATE Generate $B$ synthetic samples $D=\{\phi_g(\theta_g;z_i)\}_{i=1}^{B}$
\FOR {each synthetic data $d_i=\phi_g(\theta_g;z_i)$}
\FOR {each teacher $\phi_t^j$}
\STATE Compute loss $\mathcal{L}_T(\phi_t^j(\theta_t^j;d_i),\phi_s(\theta_s;d_i))$
\STATE Compute the gradient $g_{ij}=\frac{\partial \mathcal{L}_T}{\partial \phi_s(\theta_s;d_i)}$
\STATE Normalize the gradient $\bar{g}_{ij}=\frac{C\cdot g_{ij}}{||g||_2+e}$
\ENDFOR
\STATE Add noise $\tilde{g}_i=\frac{1}{n}(\sum\limits^{n}\limits_{i=1}\bar{g}_{ij} + \mathcal{N}(0,\sigma^2C^2I))$
\ENDFOR
\STATE Compute differentially private output of student $y_s=\phi_s(\theta_s;D)-\gamma \cdot\tilde{g}$
\STATE Compute loss $\mathcal{L}_S(\phi_s(\theta_s;D), y_s)$
\STATE Update student $\theta_s^{t+1}=\theta_s^{t}-\gamma_s \cdot\frac{\partial \mathcal{L}_S}{\partial \theta_s^{t}}$
\STATE Compute loss $\mathcal{L}_G(\phi_s(\theta_s;D))$
\STATE Update generator $\theta_g^{t+1}=\theta_g^{t}-\gamma_g \cdot\frac{\partial (\mathcal{L}_S+\mathcal{L}_G)}{\partial \theta_g^{t}}$
\ENDFOR
\STATE \textbf{return} $\theta_s$ and $\theta_g$
\end{algorithmic}
\end{algorithm}

\begin{algorithm}[!htbp]
\caption{Training Directly with Private Data}
\label{alg:TDPD}
\textbf{Input}: Examples $\mathbf{x}=\{x_i\}_{i=1}^{N}$ with labels $y=\{y_i\}_{i=1}^{N}$, number of training iterations $T$, loss function $\mathcal{L}, \mathcal{L}_S$, noise scale $\sigma$, sample size $B$, learning rate $\gamma, \gamma_s$, gradient norm bound $C$, a positive stability constant $e$\\
\begin{algorithmic}[1] 
\FOR {$t \in [T]$}
\STATE Take a random sample $B_t$ with sampling probability $B/N$
\FOR {each data $x_i$}
\STATE Compute loss $\mathcal{L}(y_i,\phi(\theta;x_i))$
\STATE Compute the gradients $g_{i}=\frac{\partial \mathcal{L}_T}{\partial \phi(\theta;x_i)}$
\STATE Normalize the gradient $\bar{g}_{i}=\frac{C\cdot g_{i}}{||g||_2+e}$
\ENDFOR
\STATE Add noise $\tilde{g}_i=\frac{1}{B_t}(\sum\limits^{B_t}\limits_{i=1}\bar{g}_{i} + \mathcal{N}(0,\sigma^2C^2I))$
\ENDFOR
\STATE Compute differentially private output of model $S_{new}=\phi_s(\theta_s;\mathbf{x})-\gamma \cdot\tilde{g}$
\STATE Compute loss $\mathcal{L}_S(\phi(\theta;\mathbf{x}), S_{new})$
\STATE Update student $\theta^{t+1}=\theta^{t}-\gamma_s \cdot\frac{\partial \mathcal{L}_S}{\partial \theta^{t}}$
\STATE \textbf{return} $\theta$
\end{algorithmic}
\end{algorithm}

\begin{figure}[!htbp]
\centering
\includegraphics[width=1.0\columnwidth]{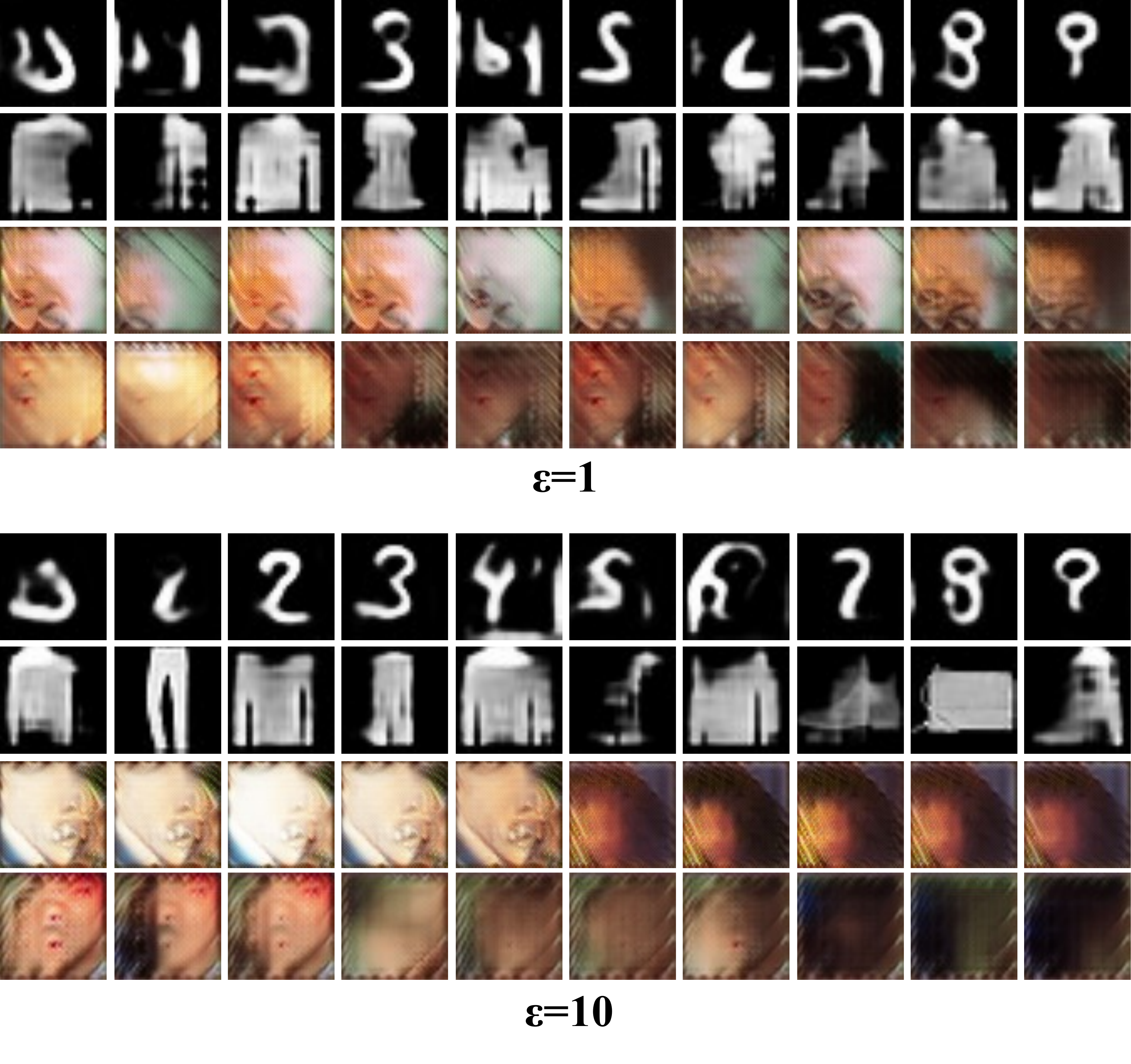}
\caption{Visualization of synthetic data under $\varepsilon=1$ and $\varepsilon=10$. MNIST, FMNIST, CelebA-G and CelebA-H in order from top to bottom.}
\label{fig:vis-mfc}
\end{figure}

\begin{figure}[!htbp]
\centering
\includegraphics[width=1.0\columnwidth]{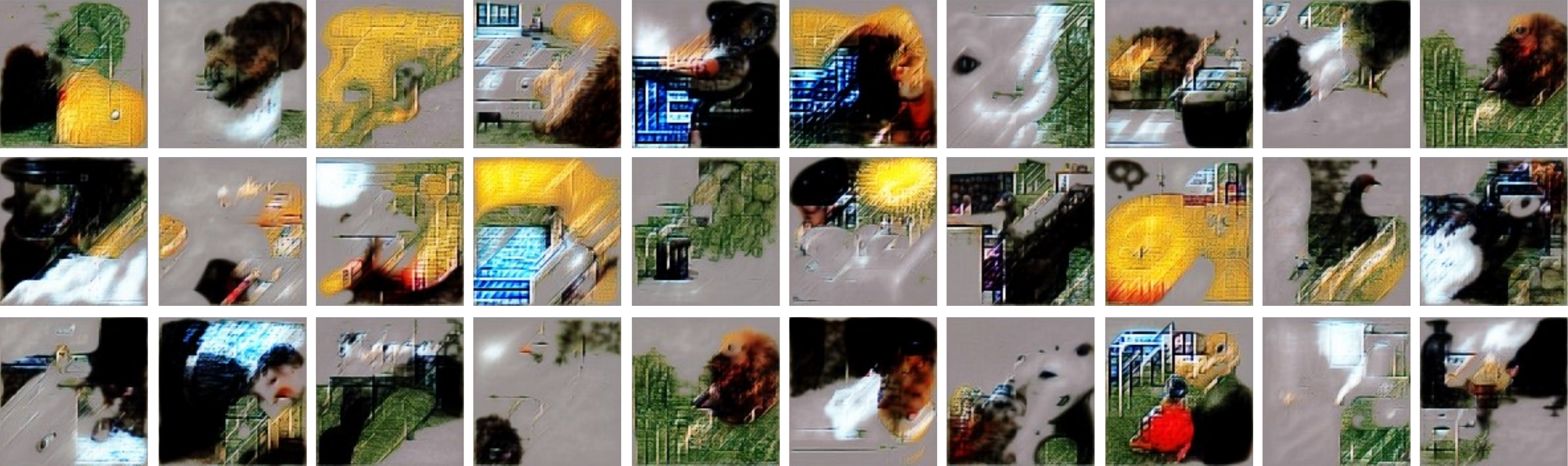}
\caption{Visualization of synthetic data for ImageNet under $\varepsilon=2$.}
\label{fig:vis-ImageNet}
\end{figure}

\subsection*{Visualization of Synthetic Data}
We visualize the differentially private synthetic data for MNIST, FMNIST and CelebA, as shown in Fig.~\ref{fig:vis-mfc}. We note that our goal is not to train a generator to generate images with high visual quality, but to generate data which can protect privacy and ensure high data utility in terms of training high-performance models. We can see that these images have a lot of noise, but enough to train useful models. As we expected, the visual quality of the synthetic images at $\varepsilon=10$ is higher compared to that at $\varepsilon=1$. However, for the high-dimensional CelebA dataset, the details of the synthetic images are still poor. We also visualize some synthetic examples for ImageNet dataset and the results are shown in Fig.~\ref{fig:vis-ImageNet}. The image also has a lot of noise. On one hand, our DP scheme makes the image noisy, and on the other hand, the data-free generator mainly learns the data distribution rather than the image details. Although the visual quality is not good, it is enough to train a useful model, which leads to interesting future direction on what machine learning models actually learn from data.

\subsection*{More Details of Convergence Analysis}
Most of our analysis process refers to~\cite{bu2022automatic}. To analyze the convergence of our DPDFD, we follow the standard assumptions in the SGD literature~\cite{allen20172,bottou2018optimization,ghadimi2013stochastic}, with an additional assumption on the gradient noise. We assume that $\mathcal{L}_T$ has a lower bound $\mathcal{L}_{*}$ and $\mathcal{L}_T$ is $\kappa$-smoothness, which can be described as follows: $\forall x,y$, there is an non-negative constant $\kappa$ such that $\mathcal{L}_T(x)-\mathcal{L}_T(y) \le \nabla \mathcal{L}_T(x)^{\top}(x-y) + \frac{\kappa}{2}||x-y||^2$. The additional assumption is that $(g_r-g)\sim \mathcal{N}(0,\zeta^2)$, where $g_r$ is the ideal gradients of $\mathcal{L}_T$ and $g$, which we compute, is an unbiased estimate of $g_r$. Because $\mathcal{L}_T$ is $\kappa$-smoothness, we have
\begin{equation}\label{eq:delta-LT}
    \begin{aligned}
    \mathcal{L}_T^{t+1}-\mathcal{L}_T^t&\leq (g_r^t)^{\top}\left(y_s^{t+1}-y_s^t\right)+\frac{\kappa}{2}||y_s^{t+1}-y_s^t||^2\\
    &=-\gamma(g_r^t)^{\top}\mathcal{A}_{C,\sigma}(g)+\frac{\kappa\gamma^2}{2}||\mathcal{A}_{C,\sigma}(g)||^2.
    \end{aligned}
\end{equation}
Given $y_s^t$, we can calculate the expectation of $\mathcal{L}_T^{t+1}-\mathcal{L}_T^t$ as follows
\begin{equation}\label{eq:Edelta-LT}
    \begin{aligned}
    &\mathbb{E}(\mathcal{L}_T^{t+1}-\mathcal{L}_T^t|y_s^t)=\\
    &-\gamma(g_r^t)^{\top}\mathbb{E}(\mathcal{A}_{C,\sigma}(g))+\frac{\kappa\gamma^2}{2}\mathbb{E}(||\mathcal{A}_{C,\sigma}(g)||^2).
    \end{aligned}
\end{equation}
Given the fact that $||\frac{C\cdot g_i}{||g||_2+e}||^2\leq C^2$.We substitute Eq.~(\ref{eq:dp}) and combine it with the Cauchy Schwartz inequality to obtain
\begin{equation}\label{eq:KS}
    \begin{aligned}
    \mathbb{E}(||\mathcal{A}_{C,\sigma}(g)||^2)\leq 2C^2+2\sigma^2C^2d,
    \end{aligned}
\end{equation}
where $d=||z||^2$, $z\sim\mathcal{N}(0,I^2)$. So we have
\begin{equation}\label{eq:Edelta-LT2}
    \begin{aligned}
    &\mathbb{E}(\mathcal{L}_T^{t+1}-\mathcal{L}_T^t|y_s^t)\leq\\
    &-\gamma C(g_r^t)^{\top}\mathbb{E}\left(\frac{g}{||g||_2+e}\right)+\kappa\gamma^2(C^2+\sigma^2C^2d).
    \end{aligned}
\end{equation}
According to the Lemma C.1 in~\cite{bu2022automatic}, we can obtain
\begin{equation}\label{eq:lemmac1}
    \begin{aligned}
    (g_r^t)^{\top}\mathbb{E}\left(\frac{g}{||g||_2+e}\right)\geq \min _{0<c \leq 1} f\left(c, r ; \frac{e}{\left\|g_r\right\|}\right) \cdot\left(\left\|g_r\right\|-\zeta / r\right),
    \end{aligned}
\end{equation}
where $f(c,r;x)=\frac{(1+rc)}{\sqrt{r^2+2rc+1}+x}+\frac{(1-rc)}{\sqrt{r^2-2rc+1}+x}$. We define $\mathcal{G}(||g_r||;r;\zeta;e)=\min_{0<c \leq 1} f\left(c, r ; \frac{e}{\left\|g_r\right\|}\right) \cdot\left(\left\|g_r\right\|-\zeta / r\right)$. According to the first assumption, it is obtained that
\begin{equation}\label{eq:l0-l}
    \begin{aligned}
    \mathcal{L}_T^0-\mathcal{L}_*&\geq\mathcal{L}_T^0-\mathbb{E}(\mathcal{L}_T)=\sum_t\mathbb{E}(\mathcal{L}_T^t-\mathcal{L}_T^{t+1})\\
    &\geq \gamma C\mathbb{E}\left(\sum_t(\mathcal{G}(||g_r^t||))\right)/2-T\kappa \gamma^2(C^2+\sigma^2C^2d).
    \end{aligned}
\end{equation}
We have 
\begin{equation}\label{eq:E(g)}
    \begin{aligned}
    \mathbb{E}\left(\frac{1}{T}\sum_t\mathcal{G}(||g_r^t||)\right)\leq \frac{2(\mathcal{L}_T^0-\mathcal{L}_*)+2T\kappa(C^2+\sigma^2C^2d)\gamma^2}{CT\gamma}.
    \end{aligned}
\end{equation}
Based on the definition of the function $\mathcal{M}$ above, we have
\begin{equation}
\begin{aligned}
    &\mathcal{G}^{-1}(x ; r, \zeta, e)=\\
    &\frac{-\frac{\zeta}{r} e+\left(r^2-1\right) \frac{\zeta}{r} x+r e x+e \sqrt{\left(\frac{\zeta}{r}\right)^2+2 \zeta x+2 e x+x^2}}{2 e-\left(r^2-1\right) x}.
\end{aligned}
\end{equation}
When $r>1$, the function $\mathcal{G}^{-1}(x)$ does not affect the monotonicity of the independent variable $x$. We define $\mathcal{F}(x)=\mathcal{G}^{-1}(x^2)$. We have
\begin{equation}
\min_{0 \leq t \leq T} \mathbb{E}\left(\left\|g_{r}^t\right\|\right) \leq \mathcal{F}\left( \sqrt{\frac{2\left(\mathcal{L}_{0}-\mathcal{L}_{*}\right)+2T\kappa \gamma^2C^2(1+\sigma^2d)}{T\gamma C}} ; \zeta, e\right).
\end{equation}
We simply set the learning rate $\gamma \propto \frac{1}{\sqrt{T}}$ and the gradients will gradually tend to 0 as $T$ increases.

\subsection*{Visualization of Attack Results}
\begin{figure}[!htbp]
\centering
\includegraphics[width=1.0\columnwidth]{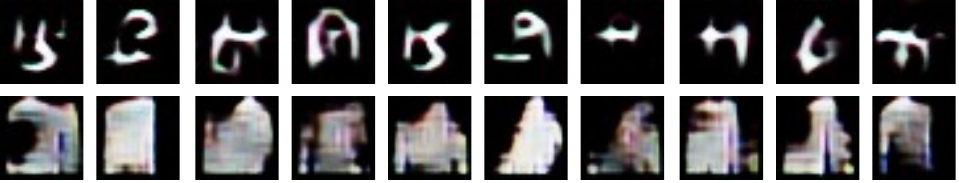}
\caption{Visualization of attack results for MNIST and FashionMNIST under $\varepsilon=1$.}
\label{fig:attack-mf}
\end{figure}

\begin{figure}[!htbp]
\centering
\includegraphics[width=1.0\columnwidth]{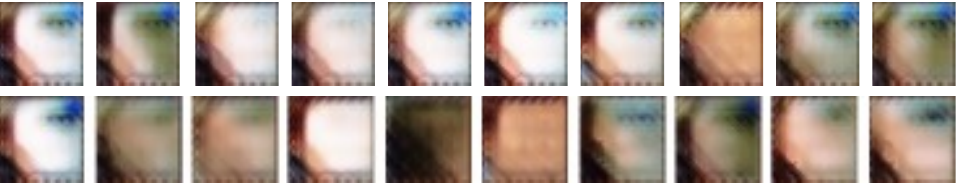}
\caption{Visualization of attack results for CelebA under $\varepsilon=2$.}
\label{fig:attack-celeba}
\end{figure}

To demonstrate the protection capability of our approach, we supplement the inversion attack~\cite{fredrikson2015ccs} experiments on the student models trained by MNIST, FashionMNIST and CelebA. Both privacy budgets $\epsilon$ of student models trained by MNIST and FashionMNIST are 1 and the attack results are shown in Fig.~\ref{fig:attack-mf}. The privacy budget of student model trained by CelebA is 2 and the attack results are shown in Fig.~\ref{fig:attack-celeba}. We can conclude from the results that attackers are unable to identify whether a particular data is in the original training data by this inversion attack.

\end{document}